\documentclass[conference]{IEEEtran}

\usepackage[utf8]{inputenc} 
\usepackage[T1]{fontenc}

\usepackage[pdftex]{graphicx}

\begin{document}
%
\title{History-Themed Games in History Education: Experiences on a Blended World History Course}

\author{\IEEEauthorblockN{Mehmet Şükrü Kuran$^{1}$, Ahmet Erden Tozoğlu$^{2}$, Cinzia Tavernari$^{3}$}
\IEEEauthorblockA{$^{1}$Department of Computer Engineering, Abdullah Gul University, Kayseri, Turkey\\
$^{2}$Department of Architecture, Abdullah Gul University, Kayseri, Turkey\\
$^{3}$Extra, Pisa, Italy\\
{E-mails:\{sukru.kuran, ahmet.tozoglu\}}@agu.edu.tr, c.tavernari@gmail.com}}

\maketitle

\begin{abstract}
In this paper we explain our experiences and observations on a blended world history course which combines classical lecture and discussion elements as well as video game sessions in which the students play strategy video games with heavy historical focus. The course, named Playing with The Past, is designed to experiment on how to integrate video games on teaching history especially in order to achieve a higher understanding of the contemporary social, political, economical, and technological context of a given era for a given nation. We ran the course four times between 2015 - 2018 with different video game titles having different historical models and observe the experiences and learning of students based on the quality of their written essays and articles. Our experiments and observations could be beneficial not only for the design of a general world history course, but also for a history course on specific periods, cultures, and nations.
\end{abstract}

\IEEEpeerreviewmaketitle

\section{Introduction}

It can be argued that one of the most difficult aspects of a history course is letting the students understand certain events, decisions, and choices for a given historical period through their contemporary lenses instead of their personal, complete modern points-of-view. Although classical teaching methodologies allow some techniques for this purpose, tools that allow experimenting with the social, political, economical, and technological norms could also be used for teaching these historical lenses. Following a "history as a process" philosophy, physical or virtual tools could be developed based on primary and secondary sources with which a student can (at least partially) "experience" a given historical era which will greatly improve their understanding of era in question.

Starting from their advent in early 1980's, many video games chose a particular historical setting (e.g., medieval Europe, crusades, warring states era Japan) as the setting of the game. Although most of these games only utilize some aspects of the historical context in their game play, some games build their game play on extensive models representing various elements of the chosen historical setting. Mostly utilized by strategy games, the focus of these models vary greatly from political, economical, technological, to militaristic. A very well-known example of such strategy games is the long-running Sid Meier's Civilization series \cite{Civ}. In this series, the player takes control of a nation starting from 4000 B.C. all the way to 2020 A.D. and experiences a variety of aspects of the nation such as technological and cultural development, expansion, military conflict, and city development. 

Considering the aforementioned "history as a process" philosophy, it can be argued that these strategy games with detailed historical models could be utilized as supportive tools in history courses \cite{Wainwright14_Teaching, Pagnotti11_Civ4, McCall16_Teaching}. By playing these games, students can freely experiment on these models to achieve a higher understanding, called an immersion process by Radetich et al., of the historical setting in question \cite{Radetich14_Using}.

We have designed an undergraduate level course named "Playing with The Past" which aims to give an introduction to world history via a blended methodology including a classical lecture and discussion (L \& D) session and game experience session. The course covers three historical ages: middle ages, early modern age, and modern age in modules. Each module starts with a L \& D session and continues with a game experience session in which the students are introduced to the mechanics of a strategy game focusing on the historical context of the module. After each module, the students are asked to play the video game on their own with certain goals and write an essay explaining their experiences and compare and contrast the historical sources on the historical setting and their individual game experiences. In addition to these essays, they are also asked to choose a nation at a certain era, play a corresponding strategy game for an extended period of time (fifteen to twenty hours) with that nation, and write down a short article focusing on their game experiences and historical sources on the same historical setting.

Over the last three years, we have experimented on using different video games in our course, namely: Sid Meier's Civilization series by Firaxis games \cite{Civ}, Total War series by the Creative Assembly \cite{TotalWar}, and Grand Strategy games by Paradox Interactive (i.e., Crusader Kings II, Europa Universalis IV, Hearts of Iron IV) \cite{CK2,EU4,HoI4}. All of these three game series have a different modeling focus. In the first two years of our lecture, we used one game from each series (i.e., Civilization IV, Crusader Kings II, and Empire Total War) and observed the students' experiences with the modeling concepts of each series. Based on our observations, the games from Grand Strategy series provided the most comprehensive experience due to their level of detail, high historical accuracy, and versatility on modeling different cultures and nations based on their specific features. Consequently, we switched to using three games in the Grand Strategy series starting from Crusader Kings II (CK2) for medieval ages, Europa Universalis IV (EU4) for early modern to industrial age, and Hearts of Iron IV (HoI4) for early - to mid 20\textsuperscript{th} century following a distinct periodization concept \cite{Koebel17_Simulating}.

In the rest of this paper, first we explain the course structure in detail, then briefly explain the main mechanics of each Grand Strategy game used in the course, and finally elaborate on our observations in combining the classical L \& D based learning and video game experience based learning in a world history course context in detail.

\section{Course Structure}

\begin{figure}[t]
	\begin{center}
    \includegraphics[width=0.45\textwidth]{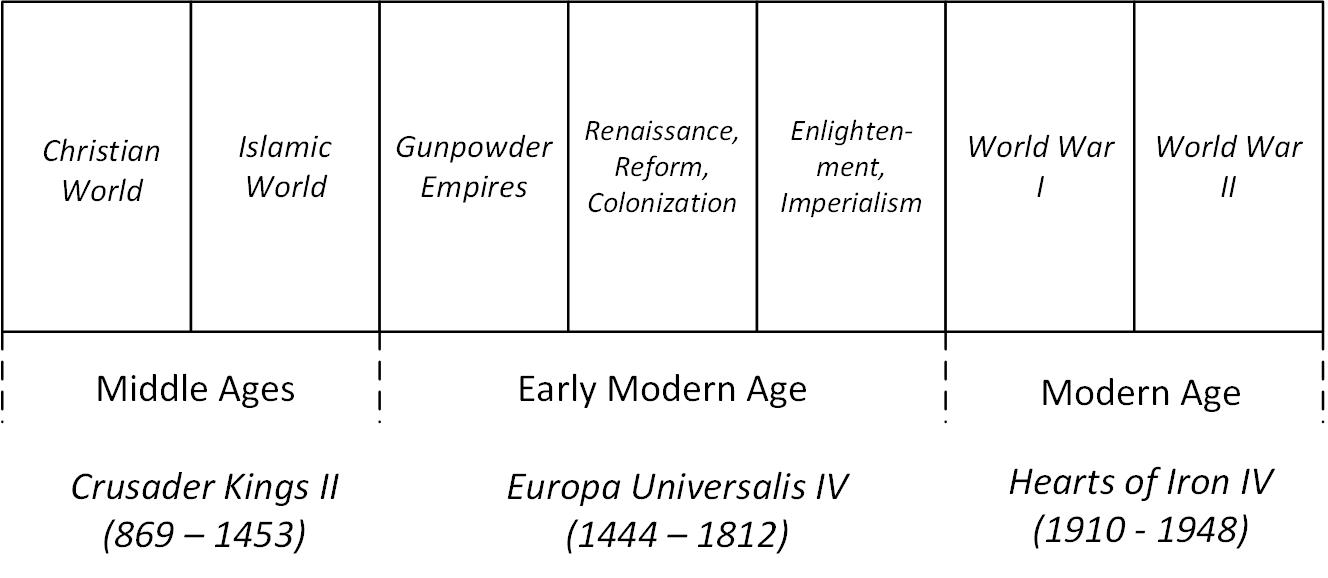}
    \caption{Structure of the ARCD301 course}
    \label{fig_arcd301_modules}
	\end{center}
\end{figure}
Our course is designed as a 5 ECTS free elective course targeted to students whose major program is not a history program. The course content is divided into three main parts as: middle ages, early modern age, and modern age (Figure~\ref{fig_arcd301_modules}). Each part is covered by  several two-week modules (i.e., 2 modules for middle ages, 3 modules for early modern age, and 2 modules for the modern age) where each module is composed of a one-week L \& D session and a one-week game experience session.

\subsection{Middle Ages Part}
In the first part, we focus on the Christian world and the Islamic world in the middle ages in two consecutive modules. In the L \& D sessions, the social and political structure of the two religious worlds have been described as well as the relationship between these two worlds (e.g., the crusades and the al-Andalus period in the Iberian peninsula). Here we underline the similarities and differences between the Feudal and the Iqta systems and the power relationships between the political and religions entities in both systems. Also, we talk about the main religious denominations in each world and the societal and political repercussions of the Great Schism in the Christian world and the Sunni - Shia split in the Islamic world.

In the game experience sessions of the first part, we introduce the game CK2 to the students and explain the mechanics on the geopolitical components, titles, religion (specifically the catholic and the sunni islam faiths), and most importantly the relationship between characters in the game. We explain the game by playing as a catholic feudal lord in Ireland as well as a sunni muslim Andalusian sultan in 1066 A.D.. Moreover, we elaborate and discuss on the modeling of the various political and social structures of the middle ages in the game. Some of these modeling aspects are apparent directly by playing the game for a short while others are indirect results of the game mechanics and can only be understood by a long immersive game experience.

\subsection{Early Modern Age Part}
In the second part, we first talk about the aftermath of the Mongol Invasion and the rise of the three muslim empires: the Ottomans, the Safevids, and the Mughals. We continue with the Reinnasance period, the Protestant Reformation, and colonization of the new world in the second module by explaining the reasons and outcomes of these periods and events. Finally, in the final module of the part we talk about the enlightenment and imperialism in the late 18\textsuperscript{th} and early 19\textsuperscript{th} centuries and the formation of the French Republic and the United States of America.

\begin{figure*}[t]
	\begin{center}
    \includegraphics[width=0.75\textwidth]{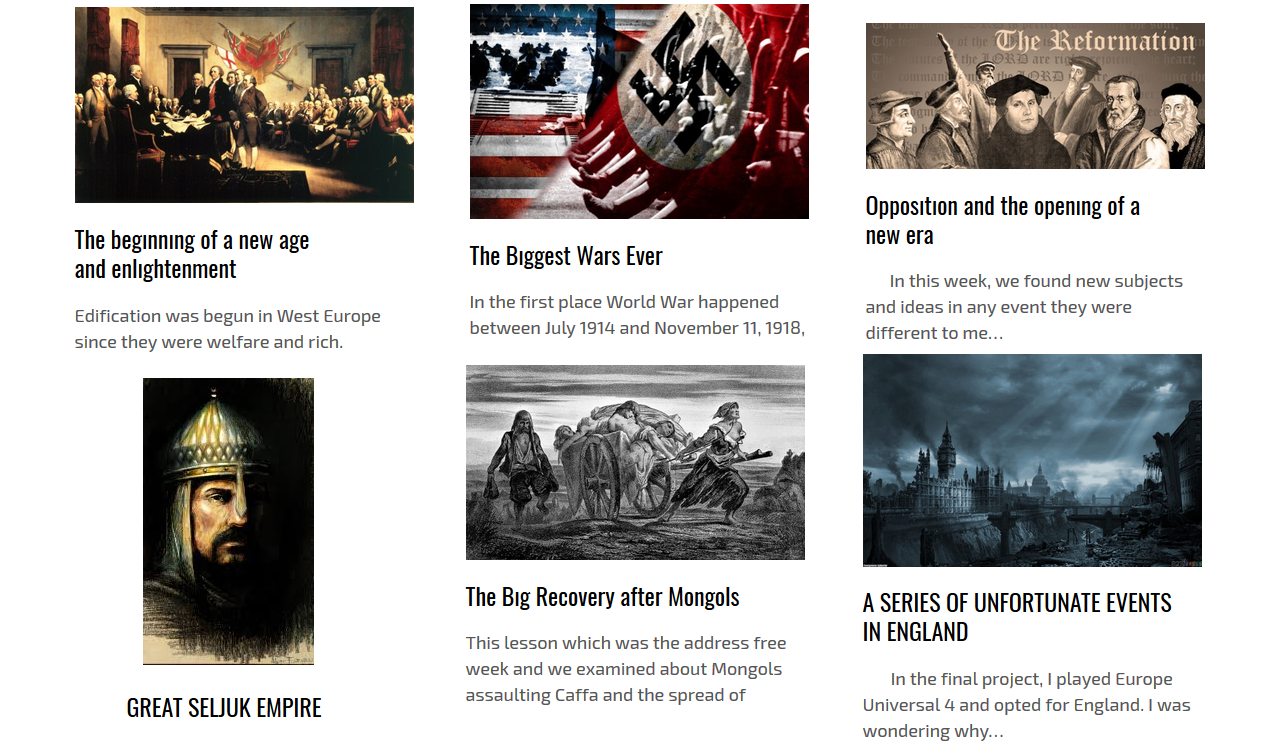}
    \caption{Example posts from the blog of our course at \textbf{playingwithpast.wordpress.com}}
    \label{fig_arcd301_blog}
	\end{center}
\end{figure*}

During the game experience sessions of this second part, we switch to EU4 and explain the differences and similarities with CK2 by focusing on the importance of Westphalian nation states instead of characters \cite{Koebel17_Simulating}, having standing armies replacing the levy system, the expanded global trade system, and technological developments. The game is explained by playing as the Ottoman Empire in 1444 A.D., Kingdom of Castille in 1492 A.D., a protestant member of the Holy Roman Empire before the Thirty Years' War, and France just before the French Revolution. Unlike CK2, nations have mission trees in EU4 which follow the historical progress of the nation in question. We elaborate on these missions by discussing on the reasons of choosing these paths in terms of game mechanics for a more immersive learning experience.

\subsection{Modern Age Part}
In the last part of the course, we start by explaining the industrial revolution and its profound implications all over the world. Then, we continue with the road to World War I, the war itself, and its results specifically the fall of the last three great empires of the early modern age. Finally, we explain the inter-war period between 1918 and 1938, World War II, and the geopolitical structure in the world after the end of World War II.

In the game experience sessions of this last part, we switch to HoI4 and talk about its game mechanics. HoI4 focuses on a very short period of time between 1936 and 1948 and tries to emulate the world around World War II. A free mod of the game, called the Great War, extends this period by letting the players start at 1910 and experience World War 1 and if they want continue forward and experience (possibly a very different) World War 2 afterwards \cite{HoI4_WW1}. For covering the time between 1812 and 1910, there is a fourth Grand Strategy game called Victoria II (V2) that focuses on the industrial revolution and pre-world war world \cite{V2}. However, the user interface of V2 is not as polished as the rest of the games and the game mechanics are much more complicated than the other games which makes V2 much more difficult for first time players to get into the game. Due to these reasons and course schedule constraints, we decided not to add a fourth game to the course with a completely different system.

\subsection{Learning Assessment}
We used three different evaluation criteria for assessing the learning of the students: weekly blog posts, a term project, and a classical midterm exam. 

After each session, the students are asked to write a 300 to 500 word long short reflection essay on the course's blog. While the blog posts after the L \& D sessions are based on the topics discussed, the blog posts after the gaming sessions are based on their game experiences after a 3 to 5 hour game session and how does the game experience correspond to the topics discussed in the previous L \& D session. In these game sessions, we limit the students' game play by offering several options to the students (e.g., in game session 2 focusing on the Islamic world in the Middle Ages, the students can either play as an Anadulician Sultan in 1066 A.D., the Fatimid Caliphate in 1167 A.D., or Sultanate of Rum in 1220 A.D.) and a different goal for each option. These blog posts are publicly available and could be found in the course's blog page \cite{ARCD301_Blog}.

After experiencing all three games described above, each student chooses a historical setting (a nation/dynasty at a certain time) on his/her own to write a 2000 to 3000 word long article on this topic by conducting a brief historical research as well as an extended gaming experience for fifteen to twenty hours as the nation/dynasty in question. In this article, the students are expected to compare and contrast the history of the nation/dynasty in question and their game experiences. 

In addition to these two assignments, we also utilize a written exam in which the students are mainly asked questions on counter factual history, some "what if scenarios". Here, they are expected to utilize their knowledge on historical sources, in-class discussions, and game experiences to come up with their own hypothesis following the aforementioned "history as a process" philosophy.

\section{Game Mechanics}

Each Grand Strategy game we've utilized in this course has extensive number of mechanics for emulating various aspects of the historical period in question (i.e., trade income, terrain features, morale of the armies). These mechanics, although not very complicated on their own, when intermixed with the other mechanics, lead to very complex interactions called emergent gameplay. In this section, we explain the key mechanics of each game while elaborating on the emergent gameplay as a result of these mechanics.

\subsection{Crusader Kings II}

CK2 has been initially designed to simulate the feudal system in Catholic Western Europe during the middle ages. With subsequent expansions, other geopolitical entities are added as playable nations such as  the Islamic world, the pre-Christian Germanic and Slavic tribal societies. As of 2018, the game map encompasses Europe, Middle East, North Africa, the Indian subcontinent, and Central Asia regions. In the game, the smallest geopolitical entity is a province which is ruled by a count or countess (or their equivalents in other cultures such as "Earl" in English culture and "Bey" in Turkic culture). Each province is further divided into several settlements as castles (governed by the nobility, a baron or baroness), cities (governed by a mayor), and bishoprics (governed by the clergy, a bishop). Several provinces next to each other form a duchy, several duchies in turn form a kingdom, and finally several kingdoms form an empire which is the biggest geopolitical entity in the game. Each one of these entities has an associated "title" which is hold by a dynasty and at a given time one character from that dynasty actually holds the title who is the designated "owner" of that title. The owner of a province is entitled to collect the tax and levy (i.e., soldiers) income from the province which constitute the main sources of economic and militaristic power in the game respectively. Lastly, each title has its own hereditary laws which dictate upon the death of the current owner who will inherit that particular title.

These geopolitical entities have a very strict hierarchical relations in the game. Since every province is part of a duchy, the owner of the duchy title is considered as the "de-jure" liege of all the owners of the provinces under that particular duchy. However, the game has a clear differentiation between a "de-jure" liege and a "de-facto" liege who may or may not be the same character. A weak de-jure duke may risk its vassals rebelling against him and form independent county-level entities. On the other hand, a strong de-jure duke can press his de-jure claim over independent count's belonging to his duchy and subjugate them after a successful war. The same system is repeated between dukes and kings.

A single character can theoretically hold as many titles as he wishes (either at the same level, or at different levels) but based on the administrative skill of the character (i.e., "demesne size") there is a limit on this number. A character exceeding his demesne size starts to get an increasingly heavy penalty on both tax and levy income which will turn his country a very ineffective political entity very fast. To remedy this issue, a character can give some of his titles to other characters (actually the dynasty of that character) which will become his vassal lords. These vassal lords are expected to take care of the province associated with that title, send some of their tax income to their lieges, and upon request lend soldiers (i.e., vassal levies) to their lieges. 

In the game, the player controls the highest ranking member of a particular dynasty. All other characters in the game are controlled by the computer and are called artificial intelligence controlled characters (AI characters). Based on many traits (e.g., genetic, social, cultural) each character has an opinion on each other character in the game which varies between -100 (complete hatred and mistrust) to 100 (complete loyalty) that affects many aspects of the game. A vassal having a bad opinion of his liege will send less amount of tax and levies to his liege and even stops sending anything if his opinion is low enough. Additionally, if there are many people having a very bad opinion of the same vassal they will start conspiring against their lord. A liege can revoke the title of a vassal if he feels threatened by a vassal but such a drastic action can have serious consequences: a) this will lower all the other vassal's opinion of the lord b) the vassal whose title is about to be revoked can declare a rebellion and go to war with his liege. Based on these mechanics, the game aims to emulate the precarious balance in the feudal system by forcing the player to have vassals and be vary of their relationships with them.

Aside from the mechanics related to the secular lords, the game has a different mechanic for the bishopric settlements. The owner of a bishopric, a bishop, has different liege, taxation, levy, and inheritance rules. A bishop has, in practice, two lieges: his feudal lord and the Pope. By default, the bishop sends his tax and levy to the Pope regardless of the geographical location of the bishopric. However, if the bishop's opinion of his feudal lord is higher than his opinion of the Pope he will send his tax and levy to his feudal lord. Additionally, in terms of inheritance, the bishop title is not a hereditary title and upon the death of a bishop the Pope assigns a new bishop to the bishopric outside the control of the feudal lord. The game allows the feudal lord to change this law (i.e., investiture law) so that he can reserve the right of appointing a bishop in his lands. Although, this alternative investiture law (i.e., free investiture) has direct benefits to the lord will in turn anger the Pope. If a lord angers the Pope too much, the Pope can excommunicate him from the Catholic faith which will allow neighboring Catholic lords to declare war on his lands and depose him. In this sense, a Catholic ruler has a second precarious relationship this time with the bishops and the Pope in particular.

Being a strategy game, the players can conduct war with other nations. However, different from many strategy games in order to wage war a country must have a valid "casus belli" in the game. There are numerous casus belli's ranging from political-based de-jure claims to religious-based conquests, excommunication wars, or even full-scale crusades or jihads. Forcing the player to require a valid casus belli to wage war is a very huge step to challenge the traditional war declaration mechanics of many strategy video games where a country (or a ruler) can declare a war as simply as whenever and wherever he or she wishes. 

\begin{figure*}[t]
	\begin{center}
    \includegraphics[width=0.65\textwidth]{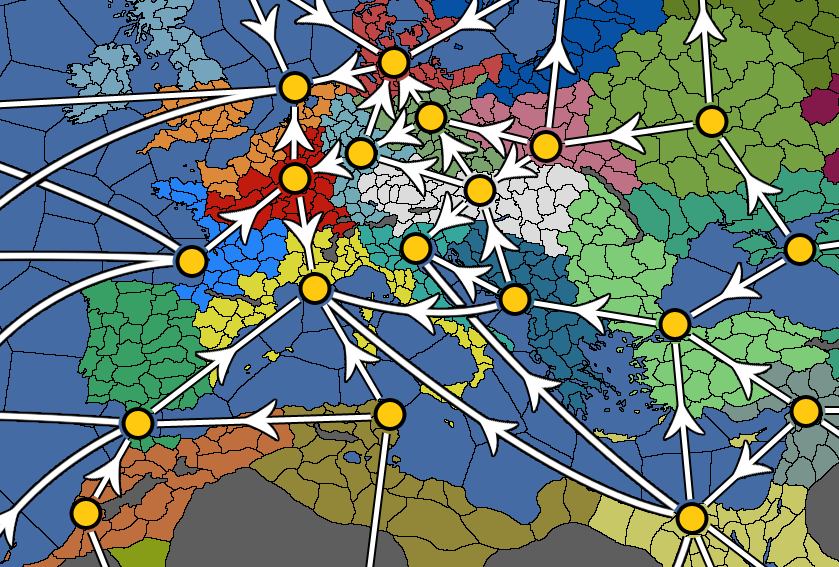}
    \caption{Trade Network of Europe, Middle East, and North Africa in Europa Universalis IV. Yellow circles specify the \textbf{trade nodes}; center of trades of the region around them given in the same color. The white lines specify the unidirectional \textbf{trade routes} which are fixed and cannot be changed throughout the game \cite{EU4_Trade}.}
    \label{fig_eu4_trade_network}
	\end{center}
\end{figure*}

Although similar to the in many aspects, the overall rules governing the muslim world is somewhat different in the game to reflect the differences between the Feudal and Iqta systems and the social implications of the Sunni Islamic faith. First of all, a muslim lord can revoke the tile of a duchy level vassal without angering his other vassals. Here the game tries to emulate a key difference of the Iqta system in which the land cannot be granted to a dynasty but instead granted to a single person and the sultan reserves the right to revoke this choice of his whenever he wishes. Additionally, since there is no centralized religious hierarchy in Sunni Islam, the bishopric settlement (called a Mosque in the game) can be governed by any character and only the sultan is the liege of the holder of the mosque title. Here, the game tries to emulate the lack of any bishopric concept in the Islamic world, it is a partially successful emulation at best.

A key difference when playing as a muslim character is the decadence mechanic. Every muslim dynasty has an associated decadence rating which ranges from 0 (no decadence, fully abiding with the Islamic laws) to 100 (complete decadence, clear well known acts against the tenets of the Islamic faith). The game assumes 25 as the normal decadence rating and gives bonuses to dynasties having a lower decadence rating (higher taxes, higher morale of the army) while punishing dynasties with high decadence rating (lower taxes, lower morale of the army). Additionally, a dynasty having at least 75 decadence risks a massive decadency revolt in which a very strong rebellious army appears and try to eradicate the whole dynasty due to their extreme decadent activities. Decadence mainly increases by male relatives of the dynasty without any titles to govern or by losing wars to non-muslim countries. On the other hand, it decreases with pious activities (i.e., going to Hajj) and successful wars against non-muslim countries. 

\subsection{Europa Universalis IV}

EU4 is designed to simulate a Westphalian nation state or a multi-cultural empire of the early modern world and roughly continues where CK2 ends. Starting from 1444 A.D., the game spans four centuries until 1812 A.D. around the start of the Industrial Revolution. The game map has been considerably expanded upon the map of CK2 by including the Americas, Africa, Asia, and Oceania. Similar to CK2, the smallest geopolitical entity in the game is again a province but this time local rulers (i.e., counts, bishops, mayors) and other dynastic characters are completely eliminated and their interactions are abstracted into an autonomy percent and a local unrest rate. The entity controlled by the player in EU4 is a "nation" having a certain government type (e.g., monarchy, republic), main culture, state religion, and an administrative ruler (e.g., king, emperor, president, etc...). Thus, the game implies that the age of character-driven political entities are over and replaced by state-level entities where the governmental traditions outweigh individual characters.

Instead of levy-based armies in CK2 which are called upon only during the war time, each country has a standing army in EU4 whether it be the time of peace or the time of war. These armies start by being composed of infantry and cavalry, but as time progresses they also incorporate mobile artillery units which eventually become a crucial part of the army in the late game in the 18\textsuperscript{th} and 19\textsuperscript{th} centuries. Unlike CK2, navies play a much more important role as able to conduct various missions such as protecting sea trade, exploration of the unknown regions, blockade rival ports and attack enemy ships. The game divides ships into two broad categories: galleys which are more suitable for inland seas such as the Mediterranean and the Baltic sea and ocean-faring ships which are much more useful in the oceans and able to explore new regions that pave the road to colonization of the new world. Using this two type of ship distinction, the game tries to separate inland sea powers (e.g., Venice, Genoa) and oceanic sea powers (e.g., Portugal, Britain, Netherlands). Similar to CK2, declaring a war mainly requires a valid casus belli which again depends on many factors. However, this time the player is also given the choice to declare a war without a valid casus belli which might be beneficial in certain conditions while seriously hurting the stability of the country as well as its international prestige.

\begin{table*}[t]
\begin{center}
\caption{Minimum world tension requirements for taking aggressive actions by ideology \cite{HoI4_WorldTension}}
  \begin{tabular}{ | c | c | c | c | c |}
    \hline
    Actions & Communist & Democratic & Fascist & Non-Aligned \\ \hline
    Guarantee Independence & -    & 25\%  & - & 40\% \\ \hline
    Send volunteers        & -    & 50\%  & - & 40\% \\ \hline
    Join faction           & -    & 80\%  & - & 40\% \\ \hline
    Justify war goal       & -    & -     & - & 50\% \\ \hline
    Declare War            & 75\% & 100\% & - & 25\% \\ \hline
    Lend-Lease             & -    & 50\%  & - & 40\% \\ \hline
    
  \end{tabular}
  \label{table:HoI4_World_Tension_Actions}
\end{center}
\end{table*}

Although CK2 includes a three-category technology system (i.e., militaristic, economical, and cultural), the individual technological advancements have somewhat of a reduced importance in the overall game. On the contrary, EU4 takes a more classical look at technological advancements and let them play a very critical role of EU4 where falling behind in technology has serious drawbacks for a country. The game divides technological advances into and three categories as administrative, diplomatic, and militaristic and 32 levels in each category. Administrative technologies increase administrative efficiency, allow newer governmental types, and increase nation-wide production. Diplomatic technologies on the other hand increase trade efficiency, colonization power, diplomatic relations, and naval power of the nation. Finally, the militaristic technology increases various aspects of the land army and plays a crucial part in the game where in a battle the army whose military technology is 2 to 3 steps ahead very swiftly delivers a crushing blow to the opposing army. 

In addition to the 32 levels of technology in each category, the game also has seven special social or technological advances called "institutions", namely in the chronological order: Feudalism, Renaissance, Colonization, Printing press, Global Trade, Manufactories, and Enlightenment. These institutions cannot be researched normally instead they appear in a semi-random province in the world at a certain year. After their initial appearance, they spread to the rest of the world whose speed is affected by various effects including conscious decision of the player and the AI. When a given institution has spread to enough number of provinces of a given country, the country embraces that institution giving it some benefits. More importantly though, technological advancement and institutions have a critical interaction as after the advent of a particular institution, each country that has not been embraced it is penalized in technological advancement speed which increases every year. This interaction forces countries not yet have been embraced institutions to lag behind in technology. Although, institutions are a very interesting mechanic, the game makes a somewhat controversial decision on them which can be considered a very "Eurocentric" choice. Most of the institutions initially appear somewhere in Europe (e.g., Renaissance always start somewhere in modern Italy, and the Printing Press always start in some province with Germanic culture with either Protestant or Reformed Christian religion). Consequently, regions geographically distant from Europe start to lag behind in technology after a while and becomes considerably far way from the European nations starting from the 16\textsuperscript{th} century.

Another key element of the game is the trade network system which becomes the main source of income for most of the countries starting from the 16\textsuperscript{th} century (Fig. \ref{fig_eu4_trade_network}). Each province is set to be part of a trade region having a single center of trade province called the trade node (e.g., the center of trade of the Iberian peninsula provinces depicted in dark green is  Sevilla Fig. \ref{fig_eu4_trade_network}). Each nation controlling some trade power in a trade node can either collects trade income from the trade node or can push that trade income through a trade route to another trade node further downstream of the network where he controls most of the trade power. A very important part of the game mechanic is the fact that all elements of this trade network are fixed and cannot be changed throughout the game. Based on the unidirectional trade routes, the game implies that eventually all trade will flow towards Western Europe. Although, this trade system tries to explain why certain nations in history (e.g., Austria - Hungary, The Ottomans) never involved in the Colonization of the Americas simply because mechanically it has never been beneficial to them, the fixed nature of the trade network can also be argued as a Eurocentric choice from a design point-of-view.

Beside these important mechanics, EU4 uses a great deal of country specific events such as the War of the Roses in England or the Iberian Wedding between Castille and Aragon kingdoms. Although, the game gives players some choices on these country - based events, generally the historically correct option is mechanically the better option. By utilizing these specific decisions, it can be argued that the game consciously tries not to diverge too much from the actual history. 

\subsection{Hearts of Iron IV}

Unlike CK2 and EU4 which cover several centuries of time each and tries to emulate the socio-political systems in the respective time periods, HoI4 specifically focuses on the WW2. Set in the time several years before WW2, players again control a country as in EU4 but this time the countries are all secular states without any associated religious mechanics but instead having a political ideology which affects many aspects of the game play. A country's ideology can be either democratic, fascist, communist, or non-aligned. Most of the diplomatic options available to a country depends on its current ideology which can also be changed either by elections, a military coup d'etat, or a civil war (e.g., Spanish Civil War). Fascist and Communist countries can declare war on other countries just like any country in EU4 if they have a legitimate claim on a foreign soil. However, Democratic and non-aligned countries must wait until a unique metric called the "world tension" to reach a certain threshold before declaring war. This metric starts at 0\% at 1936 and increases with world-wide hostile activities (e.g., Anschluss) and capitulation of countries in wars (e.g., Italio-Ethiopian war of 1930's). Even than, Democratic countries can only declare war when world tension reaches 100\% and only on countries who have contributed to the increase of the world tension (Table~\ref{table:HoI4_World_Tension_Actions}). 

Being a game focused issues related to the world war, HoI4 greatly simplifies the trade element and only focuses on production and trade of six "strategic resources", resources that are crucial for the war effort: oil, aluminum, tungsten, rubber, chromium, and steel. A country requiring these resources makes a trade agreement with a country having a surplus of the same resource. In return, the resource receiving country allocates some of its civilian factories to produce civilian goods that are freely sent to the other. If the two trading countries have a connection through land than the goods are invisibly transferred between these two countries. Otherwise, the trade becomes an overseas trade and must be carried by naval convoys.

In HoI4, armies consist of land, naval, and aerial units. Although, land units are still the backbone of the army, both naval and aerial units play key factors in any military conflict. The game greatly increases the importance of the naval units due to the aforementioned trade system and the HoI4 specific logistics system. In the game, land units fighting on a front constantly take loses on both manpower and military equipment terms. These losses are automatically replenished from the country's manpower and equipment pool as long as there is a undisturbed logistic path between the nearest region of the country to the front line that has taken these losses. While this is fairly simple for a country waging war with a neighboring country, in an overseas conflict logistic paths rely on naval convoys. In case of a war, these naval convoys (whether they are carrying strategic resources or supply to front lines) are under constant danger of being attacked and sunk by the naval and aerial units of the opposing countries and therefore must be protected by the navy and air force. This detailed logistic system underlines the new challenges in a modern military conflict and implies the requirement of completely new strategies for a successful military campaign.

\section{Observations of the Student Experiences}
In this section we explain our observations on the students' experiences in our blended world history course and underline the key advantages of using such a methodology. We base our observations on the blog posts used by the students in our course, the in-class discussions, and a survey we have conducted to the students.

\textbf{In-depth geographical knowledge:} Utilizing an interactive world map, which directly or indirectly effects many game mechanics, greatly increases the student's knowledge on world geography and its socio-political and economic implications in history. By allowing the students to interact with the map (army/navy movement, trade routes, regional development, terrain types, etc...) they internalize concepts like strategical and economic importance which in turn help them understand various key decisions taken by countries, nations, or leaders in the past.

\textbf{Increased awareness on the interaction between various societal issues:} Having an interactive tool (i.e., the games) that allows experimenting on a variety of different societal issues by taking decisions, greatly increase students' understanding of such topics. They become more aware of the complex set of interactions between economic, religious, technological, political, cultural elements and realize that in most cases none of these elements stand on their own, instead they affected each other profoundly throughout the history.

\textbf{Experience-based, immersive learning:} Most of the students report that learning history through a video game has a critical immersive component. They can look at the game, experiment on a given strategy, and get an immediate feedback from the game. Based on the feedback, they can alter their strategy and try it again to see the differences in the outcome. After a while, this trial-and-error methodology allows them to internalize the intricate interactions of the various systems in the game and lead to a more immersive learning of the subject matter.

One important side note here is most of the students who could not achieve the goals we put to them in the game-based assignments, incentivised themselves to try again by different tactics and set of decisions. Even though we explicitly stated to the students that success or failure is not important in the game-based assignments, they prefer to try again and learn how to succeed these goals. Since these additional trials become personal assignments they set up for themselves, they have a very high motivation which in turn further increases the immersiveness of their experience.

Moreover, since all the games we have used cover a long period of time in history, after a few game sessions students surmise the fact that their previous decisions affect their current situation profoundly and understand the concept of contingency in the history-as-a-process philosophy.

\textbf{Discussion over modeling decisions of the games:} 
On many occasions, either after we have explained a certain game mechanic or after a game-based assignment, students started criticizing the modeling choices of the game regarding these mechanics. We observe that having a discussion over an interactive model is much more easy to follow than discussing over a passive reading of history in a historical text. The key advantage of discussing over an interactive model is the ability to check how it works with the other game mechanics and the implications of such a model in the greater context of the whole game.

A good example of such a modeling decision is the controversial "westernization" mechanic of EU4 which has been replaced by the more versatile "institutions" mechanic explained in Section IV-B. Although, the "institutions" mechanic can still be criticized as Eurocentric, the "westernization" mechanic has been vastly more Eurocentric and led all non-western European nations to either become a backward country in the long run or change their culture and "westernize".

\textbf{Contextualizing key events and major developments:} 
Last but not least, we observe that after a couple of game sessions, students' understanding of the societal systems of the old increase considerably. Afterwards, they start to elaborate on events more from a historical perspective instead of a modern view-point. This change of perspective greatly increases their understanding of certain historical key events and developments being part of a greater process instead of thinking them as strange occurrences that happened with little connection to its contemporary surroundings.

\section{Conclusion}
In this paper, we explained our experiences and observations on a blended world history course having classical lecture and discussion sessions as well as game sessions in which students can play complex strategy games that model certain historical eras. We explain our course structure and the core mechanics of the three games we have utilized in our course: Crusader Kings II, Europa Universalis IV, and Hearts of Iron IV. Based on our observations on the students' experiences, having such a blended course in which historical video games are used as supportive tools greatly increases the students' learning in the topics in question. By experimenting with the mechanics of each game, students internalize historical knowledge as active participants instead of passive readers. Although, it requires some effort to set up such a blended world history course, we observe the gains outweigh the challenges and allow for a more deep and immersive learning experience.


\end{document}